# Gemini Planet Imager One Button Approach

Jennifer Dunn*[a], Dan Kerley[a], Leslie Saddlemyer[a], Malcolm Smith[a], Robert Wooff[a], Dmitry Savransky[b], Dave Palmer[c], Bruce Macintosh[c], Jason Weiss[d], Carlos Quiroz[e], Fredrik T. Rantakyrö[e], Stephen J. Goodsell[f]

[a]National Research Council Herzberg, 5071 W. Saanich Rd., Victoria, BC, V9E 2E7, Canada
[b]Sibley School of Mechanical and Aerospace Engineering, Cornell University, 410 Thurston Ave., Ithaca, NY 14853.
[c]Lawrence Livermore National Laboratories, 7000 East Avenue, Livermore, CA 94550.
[d]Dept. of Physics and Astronomy, Univ. of California at Los Angeles, Los Angeles, CA 90095.
[e]Gemini Observatory, Casilla 603, La Serena, Chile
[f]Gemini Observatory, 670 N. A'ohoku Place, Hilo, HI 96720 USA

## ABSTRACT

The Gemini Planet Imager (GPI) is an "extreme" adaptive optics coronagraph system that is now on the Gemini South telescope in Chile. This instrument is composed of three different systems that historically have been separate instruments. These systems are the extreme Adaptive Optics system, with deformable mirrors, including a high-order 64x64 element MEMS system; the Science Instrument, which is a near-infrared integral field spectrograph; and the Calibration system, a precision IR wavefront sensor that also holds key coronagraph components. Each system coordinates actions that require precise timing. The observatory is responsible for starting these actions and has typically done this asynchronously across independent systems. Despite this complexity we strived to provide an interface that is as close to a one-button approach as possible. This paper will describe the sequencing of these systems both internally and externally through the observatory.

## 1. INTRODUCTION

GPI is a complex instrument that contains three systems that are complicated enough to be considered instruments on their own: an Adaptive Optics system (AOC), a Calibration facility (CAL), and the Science Instrument (Integral Field Spectrograph, IFS). More details can be found in Macintosh, et al [1]. Because these systems have to work so closely together to get the best contrast ratio, they are all wrapped into the same instrument. This paper will discuss how we have hidden the complexity of the instrument behind single buttons and will largely discuss the operations during the night. Any decisions needed to be made when commands are issued are based on the configuration selected, the current state of the instrument, the light available, and the physical environment.

---

* Jennifer.Dunn@nrc-cnrc.gc.ca phone 1 250 363-6912; fax 1 250 363-0045; www.hia-iha.nrc-cnrc.gc.ca

## 2. SOFTWARE SYSTEMS

The Gemini Planet Imager is composed of four software systems (one for each of the principle systems mentioned above plus supervisory software) that were developed separately by four partnering institutions. Ultimately they work together and must appear part of the same fabric. The various parts of GPI are complex in their own right and must have a very simple interface because the Astronomer can't be expected to know the significance or effect of configuration choices in the Adaptive Optics and Science Instrument. Observations are defined by an astronomer using the Gemini Observing Tool (OT) for all facility instruments. These observation configurations are considered to be "observing recipes". More details on the configuration in Gemini are detail in [2]. Commands in GPI are hierarchical and pass from the Gemini interface to the GPI Instrument Sequencer. The role of the Instrument Sequencer is to provide an interface where the majority of the sequencing is done within the instrument. From the perspective of the observatory, it is essentially a black box. The options seen in the Observing Tool by the astronomer are geared towards the types of observations that the observer wants to do rather than dealing with the nitty-gritty details of how to get GPI to do the type of observation selected. Gemini provides a layer of translation between the Observing Tool and the commands that are actually sent to the GPI Instrument Sequencer.

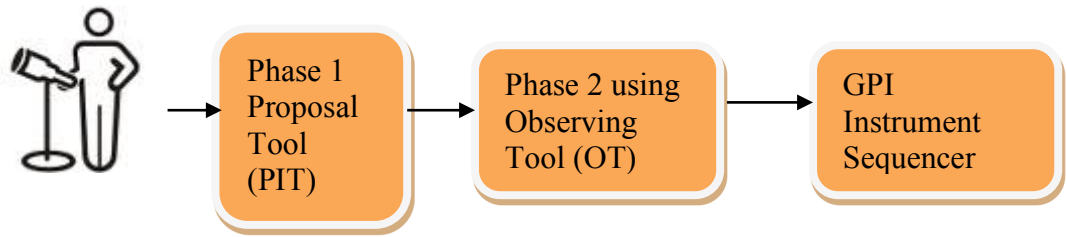

**Figure 1 Observation Transition**

A single setting for GPI can result in multiple functions and/or settings within the instrument. This paper will highlight those areas where a significant amount of sequencing has occurred. To understand what sequencing is performed it is important to understand the components inside GPI. Figure 2 shows some of the significant GPI components:
- Atmospheric Dispersion Corrector (ADC)
- DM1 – woofer
- DM2 – MEMS tweeter
- Pupil Plane Mask (PPMs) (Apodizers in baseline)
- Tracking assemblies (fold mirror, AO Wavefront Sensor Pointing and Centering Mirrors, CAL Pointing and Centering Mirrors )
- Adaptive Optics Wavefront Sensor (AOWFS )
- Focal Plane Mask (FPM)
- Low Order WFS (LOWFS)
- High Order WFS (HOWFS)
- Polarization modulator (in the CAL) and Polarizing Beamsplitter (in the IFS)
- Lyot Mask
- Hawaii 2RG Science Instrument

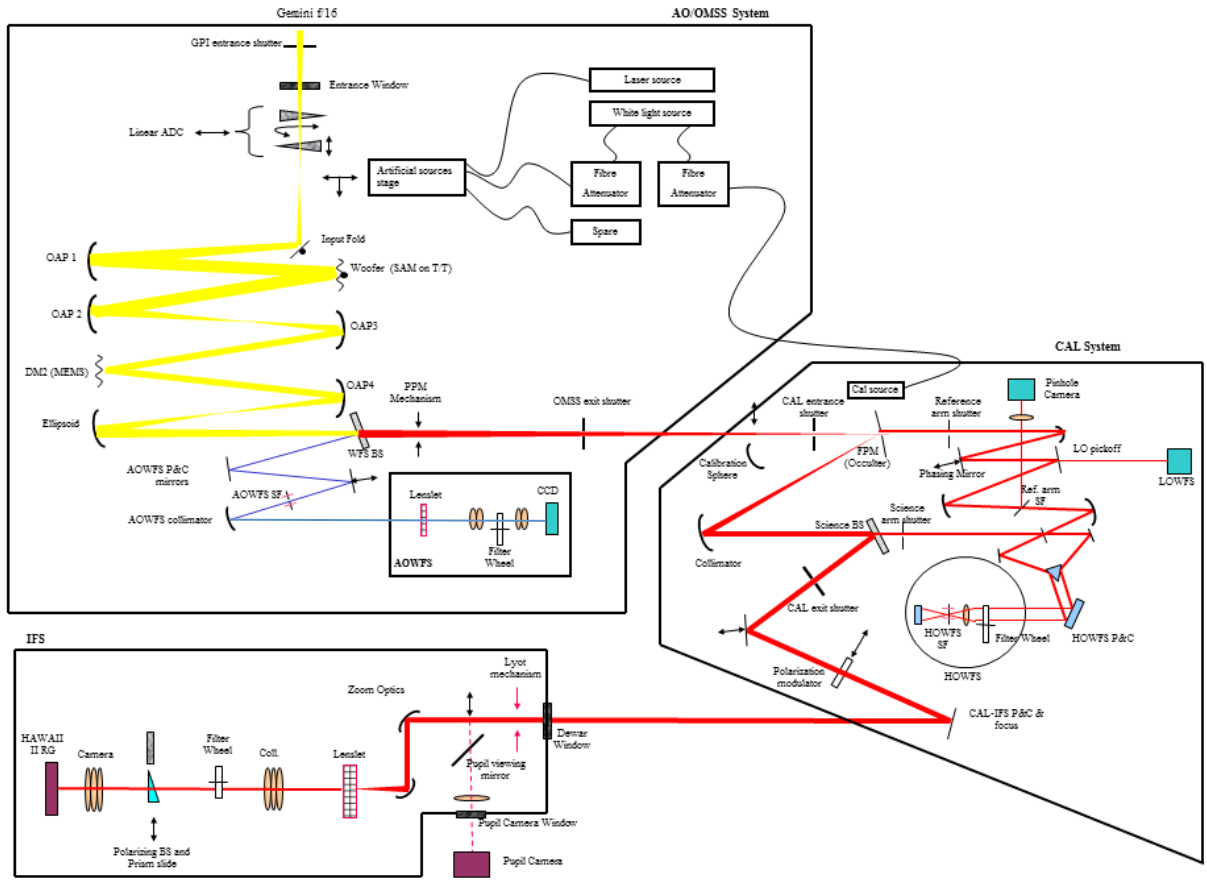

**Figure 2 GPI System Sketch**

The heart of the Adaptive Optics (AOC) system is a pair of deformable mirrors. The first is a high-stroke 11 actuator diameter piezo DM, called the "woofer" (the active optical area is 9 actuators across, but the mirror is 11 actuators across to give us a ring of slave actuators; and, actually, the actuator pattern is round, not an 11x11 square), which is mounted onto a tip/tilt stage to provide fast T/T response corrections. The woofer corrects high-amplitude but relatively low spatial frequency wavefront errors so that the second DM, a high-order 64x64 actuator MEMS device, called the "tweeter", can provide finer correction.

The Calibration system (CAL) is used to sense wavefront errors at the coronagraph focal plane mask (FPM). Time-averaged error signals are fed back to provide corrections to the pointing and centering optics that maintain co-alignment between the FPM on the CAL and the AOC system. The CAL system is also capable of sending high-order wavefront corrections in the form of reference wavefront offsets to the AOC, but this functionality is not currently in use.

The Science Instrument is an integral field spectrograph (IFS) utilizing a lenslet array in the field focal plane. It employs a 2048×2048 pixel HAWAII-2RG with a field of view of 2.8 arcseconds and 14 milliarcsecond spatial sampling.

## 2.1 GPI night operations

GPI night operations are divided into four categories:
- Startup Sequence: startup at the beginning of the night
- Apply Correction: correction using the Adaptive Optics and Calibration system
- Observation Sequence: performed on a target.
- Parking Sequence: parking the system at the end of the night

Not included in this paper are the daytime calibrations, that are performed frequently and do not require a dark sky, and the engineering tasks, which are sequences that require offline data processing or are done infrequently, for example generating pointing models or taking pupil images for alignment.

## 3. STARTUP SEQUENCE

GPI has both a static configuration that is maintained over the course of an observation and a dynamic one that will adapt during an observation. The Startup sequences are done once each night to prepare GPI and include all tasks required to prepare the instrument for an observation. This may be from a cold start (powered off) or from a "parked" configuration from the previous evening. Re-initialization of all software components and a movement to initial positions is performed at the beginning of the night. Then the automated alignment of GPI occurs using a collection of IDL scripts that move components, make measurements, process data, and adjust accordingly. These alignment details are presented in Savransky, et al [3] and [4].

## 4. APPLY CORRECTION SEQUENCE

The Apply Correction sequences setup and enable the Adaptive Optics and Calibration systems. This is achieved from the Gemini software side by applying the configuration for the observation that will perform the "Alignment and Calibration" of GPI.

The Adaptive Optics (AOC) system configuration includes the estimation of the star I-Magnitude. This will set the wavefront sensor camera rate, the AO filter selection and the final size of the spatial filter size to be used when the AOC system has settled.

Configuration of this system involves the H-magnitude of the central light that is presented to the CAL system, as the CAL system can only see H band light.

When a new target is acquired it is important to perform internal calibrations. This is where a significant portion of the sequencing occurs. The first step in the internal calibrations sequence is to save every setting so that it can be restored when calibration has completed and then perform alignment and calibration of both the AOC and CAL system. The second part is to reset the system back to be on sky.

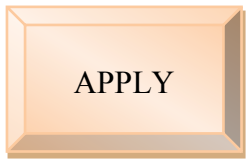

## 4.1 Align and Calibration

When observing a science object calibrations are performed just prior to observing to ensure accurate alignment and so that they are done at the representative gravity vectors and current ambient environmental conditions. Once the instrument is at the correct telescope orientation, the first step is to close off the star light and use the internal calibration light source. The light source is used to avoid atmospheric aberrations when aligning some of the more critical components.

This involves re-configuring both the AOC and CAL system to be able to use the internal light source, including re-taking the AOC system darks. This puts the tracking assemblies into a mode where it will take into account the current temperature, selected filters, the atmospheric dispersion corrector state, and gravity flexure effects.

At this point the woofer and tweeter AO loops are briefly closed which will cause the deformable mirrors and tip/tilt stage to be controlled, resulting in a flat wavefront. Then the alignment between the AOC tweeter and wavefront sensor lenslets is measured and adjustments are made.

In the next step, the full self-calibration will occur between the AOC and CAL system. This is the same function that occurs when on the sky, called "GUIDING". In this case however we guide on the internal light source. Guiding has 11 steps and is detailed in the Section 4.2.

If desired, before the next step of the alignment and calibration procedure, the IR light source is deployed and an image taken for calibration purposes.

The final step is to stop the "GUIDING", leave all offsets as they are. At this point, the system is well-aligned and will be available for guiding on a star.

## 4.2 Guiding

After the Alignment and Calibration has occurred on target then the Telescope Operator will issue one command, "GUIDE".

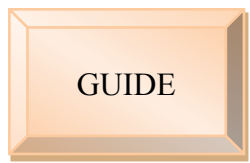

Guiding starts with re-setting offsets where appropriate and re-starting all tracking mechanisms in the tracking assemblies. Tracking systems in GPI monitor and respond to changes in elevation, temperature, etc.

The CAL system reports corrections for H band which could be different than the band used by the science instrument; in this case, an offset bias must be applied to the mirrors that guide the light into the science instrument. Depending on the observation in progress, the required offset bias will have different values. These CAL bias offsets are adjusted during this phase.

Guiding will also perform housekeeping for the AOC system such as zeroing-out offsets, clearing past observation settings, etc.

The spatial filter is a spatial anti-aliasing filter for the AO wavefront sensor and is open for the initial acquisition, when the WFS detects a seeing-limited star (Poyneer, et al [4]).

Initially the AOC's tip/tilt and woofer system are put into closed loop. Corrections are fed to the woofer and tip/tilt stage as a result of the fast corrections determined by the real-time calculations.

After the low order loop has been closed, the MEMS tweeter DM is put into closed loop. The AO system also offloads to the input fold mirror to allow for the adjustment of the centering of the Gemini pupil. At this time, offloading to the Gemini primary mirror is sent using Zernikes coefficients but currently are not applied by Gemini; and offloading to the secondary mirror is through a dedicated reflective memory card.

At this point the AOC system is offloading not only to the DM's and tip/tilt stage, but also the pointing and centering pair of mirrors that guide light into the AO wavefront sensor, and another pair of pointing and centering mirrors that guide light from the output of the CAL system into the Science Instrument. After these mirror pairs have settled, the spatial filter is stepped down to a cutoff point at a single wavelength. Then the AOC system is ready to start optimizing.

Next the CAL system is engaged. The light from the AOC system is passed through a transmissive apodizer pupil plane mask (Soummer, et al [6]) and brought to a focus at a focal plane mask where the central core of the light is removed and sent into the CAL system to be measured. Fifty percent of the on-axis light is used by a Shack-Hartmann Low-order wavefront sensor (LOWFS) which senses the average near-IR (1.6 micron) wavefront at the focal plane mask, particularly tip and tilt. The remainder, together with a portion of the off-axis light, is used to feed a high-order interferometer (HOWFS, Wallace, et al [7]), which is not currently operational. At the start of every GUIDE on the internal light source, the CAL system centers the starlight onto the coronagraphic focal plane mask (FPM). It does this by dithering in each of four directions, and steering the light until the beam evenly clips the edge of the FPM in each direction. Steering is accomplished by sending commands to the pointing and centering pair feeding the AOWFS - this displaces the light on the AOWFS, which responds by commanding the tip/tilt mirror to steer the beam back onto the AOWFS boresight, causing the beam to also move at the CAL.

Light is aligned well at this point and both the low and high order correcting has started. The tip and tilt that are measured with the LOWFS are offloaded into the mirrors that steer light into the AO WFS, which subsequently causes movement of the light to the CAL. Currently the offloading of phase data from the CAL High Order Wavefront Sensor to the AOC system is not implemented.

## 5. OBSERVATION SEQUENCE

An observation can include multiple exposures and is predefined from an observing recipe executed by Gemini.

This will include configuration of GPI for the observation, acquiring the new target, initiating the "GUIDE" procedure which will start applying correction to the wavefront, and taking a series of exposures with the science camera, then saving the science and calibration data**.** Observations are further divided into the following:
- **Configure**: Configure GPI mechanisms for a new target configuration (may be the same target, just a different configuration).

- **Observe**: Take one or more science exposures and perform any required tasks during readout and while the data is written.

## 5.1 Configuring

A set of configuration information is transmitted from Gemini to the GPI Instrument Sequencer between each observation, including:
- Set observation mode. This is a pre-configured list of modes that will command many parts of the instrument.
- ADC
- Offset of the Science Object
- Source/Shutters
- Disperser and Polarizer
- Configure exposure

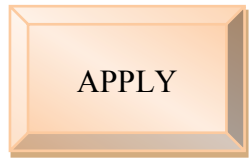

The observation mode specifies the band, and will physically select the appropriate filters. It will also select the orientation for the observation which is important for GPI because there are eight Lyot masks in GPI that also mask the secondary support spiders. This will also allow for the selection of the apodizer mask.

When observing at orientations angle not near zenith the ADC is inserted and tracks the current zenith and cassegrain rotator angles.

GPI has an internal light source which provides a convenient means to perform calibration and also has control of five shutters, which are generally not opened until an observation is performed. The exception is that appropriate shutters are opened during calibrations.

GPI has a disperser and a Wollaston prism that can be put in the light path to maximize the separation on the detector between the polarized spots of adjacent lenslets. Typically in a polarization observation sequence the angle of dispersion is modified for each exposure taken in an Observation sequence.

The Science Instrument IFS detector can be configured to select a rectangular sub region (theoretically providing a faster frame rate, though detector architecture issues limit the utility of this), the sampling mode (CDS, MCDS, up-the-ramp), the number of sampling groups (in an up-the-ramp sequence) and the number of reads per group, and the number of co-additions. The default is to take full frame images with a sampling mode of up-the-ramp.

## 5.2 Observing

An observation involves
- the opening of shutters,
- the starting of an exposure,

- the reading out of the detector,
- any real-time data processing such as reference pixel subtraction, sampling arithmetic, and combining co-adds,
- writing a FITS file.

This is achieved by Gemini sending 1 command, "OBSERVE".

Once raw FITS files are written, they are automatically detected by the GPI data pipeline's Quicklook mode, and automatically reduced using the most recent wavelength solutions, dark frames, and other calibrations, then displayed on the GPITV Quicklook (Perrin et al 2014 [8]).

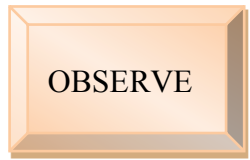

## 6. SEQUENCING PARKING

When the instrument is not going to be used for a time then the parking sequence is issued. This involves putting all devices into a state where the system is in a safe position until the next observation night. This is also the state if you need to power off the instrument.

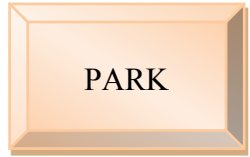

## 7. SUMMARY

The one-button approach is advisable when there is nothing to gain by exposing users to multiple steps that require decisions based on knowledge about how the system is best used; indeed, complex systems are made more reliable by having the system make these decisions on behalf of the astronomer. GPI presents the following "one-button" commands:

- Startup: prepare instrument at beginning of night for use.
- Apply Align and Calibration: internally align and calibrate all systems at the given conditions.
- Guide: track object once aligned and calibrated.
- Observe: enact an observation configuration and execute an observation.
- Park: put the system into a state that is safe for a long period of in-activity.

## 8. ACKNOWLEDGEMENTS

The authors would like to acknowledge the contributions of the entire GPI integration, testing, commissioning and science teams as well as the invaluable assistance of the Gemini South telescope operations staff. Jennifer would also like to acknowledge the support of her family: Rodger, Jade, Marina, and Christopher.